\documentclass{amsart}

\newcommand{\Reals}{{\mathbb{R}}}
\newcommand{\Cmplx}{{\mathbb{C}}}
\newcommand{\Ints}{{\mathbb{Z}}}
\DeclareMathOperator*{\supp}{supp}
\let\Re=\undefined\DeclareMathOperator*{\Re}{Re}
\let\Im=\undefined\DeclareMathOperator*{\Im}{Im}

\newtheorem{thm}{Theorem}[section]
\newtheorem{prop}[thm]{Proposition}

\theoremstyle{definition}

\newtheorem{Example}[thm]{Example}

\begin{document}

\title{Absence of reflection as a function of the coupling constant}
\author{Rowan Killip}
\address{Rowan Killip\\
         UCLA Mathematics Department\\
         Box 951555\\
         Los Angeles, CA 90095}
\email{killip@math.ucla.edu}
\thanks{The first author was supported in part by NSF grant DMS-0401277 and a Sloan Foundation Fellowship.}
\author{Robert Sims}
\address{Robert Sims\\
         UCD Department of Mathematics\\
         One Shields Avenue \\
         Davis, CA 95616}
\email{rjsims@math.ucdavis.edu}

\begin{abstract}
We consider solutions of the one-dimensional equation $-u'' +(Q+ \lambda V) u = 0$
where $Q: \mathbb{R} \to \mathbb{R}$ is locally integrable, $V :
\mathbb{R} \to \mathbb{R}$ is integrable with supp$(V) \subset [0,1]$,
and $\lambda \in \mathbb{R}$ is a coupling constant. Given a family of
solutions $\{ u_{\lambda} \}_{ \lambda \in \mathbb{R}}$ which satisfy
$u_{\lambda}(x) = u_0(x)$ for all $x<0$, we prove that the zeros of
$b( \lambda) := W[u_0, u_{\lambda}]$, the Wronskian of $u_0$ and
$u_{\lambda}$, form a discrete set unless $V \equiv 0$. Setting $Q(x)
:= -E$, one sees that a particular consequence of this result may be
stated as: if the fixed energy scattering experiment $-u'' + \lambda V u = Eu$ gives
rise to a reflection coefficient which vanishes on a set of couplings
with an accumulation point, then $V \equiv 0$.
\end{abstract}

\date{\today}

\maketitle


\section{Introduction}

The purpose of this short note is to prove a result concerning perturbations of one-dimensional
Schr\"odinger-type equations. Let $Q:\Reals\to\Reals$ be locally integrable and let
$u_0:\Reals\to\Cmplx$ be a solution of
\begin{equation} \label{Eq1}
-u_0''(x) + Q(x) u_0(x) = 0
\end{equation}
that is not identically zero.
The perturbation is a second (real-valued) potential, $V \in
L^1(\Reals)$ of compact support which, for simplicity,
we assume satisfies $\supp(V)\subseteq[0,1]$.  Define
$u_\lambda$ as the solution of
\begin{equation} \label{Eq2}
-u_\lambda''(x) + Q(x) u_\lambda(x) + \lambda V(x) u_\lambda(x) = 0
\end{equation}
that obeys $u_\lambda(x)=u_0(x)$ for all $x<0$.
The parameter $\lambda$ is known as the coupling constant.
In the problems of interest to us, it is real; however, we will allow it to vary over the complex plane as
this does not affect our results.

The question we wish to discuss is the following: for how many values of $\lambda$ is it possible that
$u_\lambda(x)$ is a multiple of $u_0(x)$ in the region $x>1$?  An equivalent formulation is to study the zeros
of the Wronskian between $u_0$ and $u_\lambda$:
\begin{equation} \label{Eq3}
b(\lambda) : = W[u_0,u_\lambda](x) = u_0'(x)u_\lambda(x) - u_0(x)u_\lambda'(x)
\end{equation}
for any $x>1$.

\begin{thm}\label{T1}
The zeros of $b(\lambda)$ form a discrete set unless  $V\equiv 0$.
\end{thm}

Furthermore, if $b$ has infinitely many zeros, they must approach infinity rather rapidly:

\begin{thm}\label{T2}
If $V\not\equiv 0$, then the number of roots of $b(\lambda)=0$ in the disk $|\lambda|\leq r$
(counted by multiplicity) is $O(r^{1/2})$ as $r\to\infty$.
\end{thm}

When $V$ is sign definite and $Q\equiv 0$, Chadan has shown that the zeros of $b(\lambda)$ determine
$V$ precisely in a slightly different scenario; see \cite{Chadan}.  A Liouville transformation is used
to reduce the problem the to well-known uniqueness problem in the energy parameter.

The proof of Theorems~1 and~2 consist of using some classical analysis to reduce the problem
to one treated by Stolz, \cite{S}. (We repeat his solution for the convenience of the reader.) As with us,
Stolz was interested in such matters for their relation to localization for the Anderson model.
Before elaborating this point, let us first explain the connection of our results to scattering theory.

Consider the time-independent Schr\"odinger equation with potential $q(x)$:
\begin{equation}\label{TDSE}
- \psi''(x) + q(x)\psi(x) = E\psi(x),
\end{equation}
which describes the wave function of a quantum particle with energy $E$.
For certain choices of $q$ and $E$, this equation admits a solution, $u_0$, that corresponds to the particle travelling
from right to left (the complex conjugate solution represents left-to-right motion).  For example,
when $q\equiv0$ and $E=k^2$ with $k>0$, we have $\psi(x)=e^{-ikx}$.
The Floquet-Bloch waves form another example when $q$ is periodic.

If a perturbation $\lambda V$ is introduced, this may cause the particle to be reflected back with non-zero probability.
This situation can be analyzed by looking at the solution $u_\lambda$ of \eqref{Eq2} with $Q(x)=q(x)-E$.  For $x>1$,
we may write $u_\lambda(x)=\alpha u_0(x) + \beta \bar u_0(x)$ for some complex numbers $\alpha$ and $\beta$.
In this way, we obtain a formula for the probability of reflection: $|\beta/\alpha|^2$. We also see that there will
be no reflection if and only if $u_\lambda$ and $u_0$ are linearly dependent on the interval $[0,1]$.  That is, there
will be no reflection if and only if $b(\lambda)=0$.

Our interest in this question stems from its relevance to the
one-dimensional Anderson model.  Consider solutions of (\ref{Eq2})
in the situation where the coupling constant $\lambda$ is a random
variable. As was first observed in the context of the random dimer model
\cite{DatKun, BiGer}, it is possible for the localization length to diverge (i.e., Lyapunov exponent vanish)
at a fixed energy. In particular, this happens at positive energies when the reflection coefficient
associated to the single site potential vanishes almost surely for a particular choice of random
coupling constant.  The presence of these exceptional energies leads to interesting transport phenomena (see
\cite{DLS,JSS} and references therein), while away from them one may prove exponential and dynamical
localization (see \cite{DSS, DSS2}). A consequence of Theorem~\ref{T1} is that these divergences cannot occur
unless the distribution of the couplings is purely discrete.

We assumed from the very beginning that $V$ was an $L^1$ function.  The theorems above are false if $V$
is permitted to be a measure as the following example shows; see \cite{DSS2}
for an analogous example in the discrete setting.

\begin{Example}
Consider $Q(x)=-k^2$ and $u_0(x)=e^{ikx}$.  For  $V(x)= \delta(x) - \delta(x-1)$, a simple calculation reveals
\begin{align}
b(\lambda) &= \lambda \left( 1+ \tfrac{\lambda}{2ik}\right) \left(e^{2ik}-1\right).
\end{align}
Thus we see that $b\equiv 0$ whenever $k=n\pi$, $n\in\Ints$. (The case $k=0$ follows by a limiting argument.)
The same is true for any other choice of $u_0$.  This is particularly evident when $u_0=\sin(n\pi x)$ for in this case,
$u_\lambda(x)=\sin(n\pi x)$, which vanishes on the support of $V$.
\end{Example}

One may ask for analogues of the Theorems given above when $V$ is not of compact support.  In this case, one would
define $u_\lambda(x)$ by the constraint $u_\lambda(x)-u_0(x)\to 0$ as $x\to-\infty$ and define $b(\lambda)$
as the limit of $W[u_0,u_\lambda](x)$ as $x\to+\infty$.  Naturally, one would need to ensure that $V$ decays fast
enough so that these limits exist; moreover the choice of decay rate cannot be made without knowledge
of the behavior of $Q$ at infinity.  We haven chosen not to pursue
this matter.

\section{Proofs}

Let us choose a solution $v_0$ of \eqref{Eq1}, linearly independent of
$u_0$, normalized by the requirement $W[u_0,v_0]\equiv 1$.
As $u_\lambda(x)$ is also a solution of \eqref{Eq1} in the region $x>1$, we may write
$$
u_\lambda(x)=a(\lambda) u_0(x) + b(\lambda) v_0(x) \qquad\text{for all $x>1$;}
$$
moreover, by computing Wronskians, we see that $b(\lambda)$ is the same function defined in \eqref{Eq3}.

First we show that $a(\lambda)$ and $b(\lambda)$ are analytic functions of order one-half.  This allows us to
deduce Theorem~\ref{T2} from Theorem~\ref{T1}.  Moreover, it shows that if the zeros of $b$ are not
discrete, then $b$ must vanish identically.  These two applications require us only to treat $b$; however
when $u_0$ is complex valued, we will need to invoke properties of $a$ when we prove Theorem~\ref{T1}.

\begin{prop}\label{P1}
The functions $a(\lambda)$ and $b(\lambda)$ are entire and obey
\begin{equation}\label{ab_bound}
|a(\lambda)| \leq C \exp\bigl\{c|\lambda|^{1/2}\bigr\},\qquad
|b(\lambda)| \leq C \exp\bigl\{c|\lambda|^{1/2}\bigr\}
\end{equation}
for some positive constants $c$ and $C$, which depend on $u_0$ and $\|V\|_{L^1}$.
\end{prop}

\begin{proof}
As in the preceding paragraphs, let $v_0$ be a solution of
\eqref{Eq1} which satisfies $W[u_0,v_0]\equiv1$.
We define
$$
K(x,t) = [ u_0(x)v_0(t) - v_0(x)u_0(t) ] V(t)
$$
so $u_\lambda(x)$ can be constructed as the solution of the Volterra integral equation
\begin{equation}\label{VoltE}
u_\lambda(x) = u_0(x) + \lambda \int_{0}^x K(x,t) u_\lambda(t) \,dt
\end{equation}
acting on $C^0([0,\infty))$.  While this equation arises naturally from variation of parameters, it is quicker
to check it by differentiating both sides twice.  The key observations are
$$
K(x,x)=0,\ \ \frac{\partial K}{\partial x}(x,x)=V(x),
\ \ \text{and}\ \ \frac{\partial^2 K}{\partial x^2}(x,t)=Q(x)K(x,t).
$$

One can solve \eqref{VoltE} by repeated substitution,
which gives rise to an infinite series for $u_\lambda$:
\begin{equation} \label{u_series}
u_\lambda(x) = u_0(x) + \sum_{n=1}^\infty \lambda^n
\int\!\!\cdots\!\!\int K(x,t_1)\cdots K(t_{n-1},t_n) u_0(t_n) \,dt_1\cdots dt_n
\end{equation}
where integration takes place over the region $0<t_n<\cdots<t_1<x$.
Convergence of this series is a well-known property of Volterra operators (cf. \cite[\S 36]{Yoshida})
and can be deduced from the estimates below.

{}From \eqref{u_series} we obtain power series for $a$ and $b$:
\begin{align*}
a(\lambda) &= 1 + \sum_{n=1}^\infty \lambda^n \int\!\!\cdots\!\!\int_{\Delta_n} v_0(t_1) V(t_1) K(t_1,t_2) \cdots
    K(t_{n-1},t_n) u_0(t_n)\,dt_1\cdots dt_n \\
b(\lambda) &= - \sum_{n=1}^\infty \lambda^n \int\!\!\cdots\!\!\int_{\Delta_n} u_0(t_1) V(t_1) K(t_1,t_2) \cdots
    K(t_{n-1},t_n) u_0(t_n)\,dt_1\cdots dt_n
\end{align*}
where $\Delta_n$ is the simplex $0<t_n<\cdots<t_1<1$.  Our bound on the size of these functions will follow by
estimating the individual terms in these series.  We only give details for $b(\lambda)$ because the argument
for $a(\lambda)$ is almost identical.

As $u_0$ and $v_0$ are $C^1$, one may choose a constant $M$ so that
$$
|u_0(t)| \leq M \quad\text{and}\quad \bigl| K(s,t) \bigr| \leq M\, |s-t| \,|V(t)| \qquad \forall t,s\in[0,1].
$$
Secondly, by the arithmetic/geometric mean inequality,
$$
\prod_{j=1}^{n-1} |t_j-t_{j+1}| \leq (n-1)^{-(n-1)}.
$$
Combining these two observations, we can deduce
\begin{align*}
|b(\lambda)| &\leq \sum_{n=1}^\infty  \frac{|\lambda|^n M^{n+1}}{(n-1)^{n-1}} \int\!\!\cdots\!\!\int_{\Delta_n}
    |V(t_1)| \cdots |V(t_n)| \,dt_1\cdots dt_n \\
&\leq \sum_{n=1}^\infty  \frac{|\lambda|^n M^{n+1}}{n!(n-1)^{n-1}} \|V\|_{L^1}^n
\end{align*}
The resulting bound on $|b(\lambda)|$ can now be deduced either through the properties of the Bessel function
$$
\sum_{n=0}^\infty \frac{r^{2n}}{(n!)^2} = I_0(2r) = \tfrac{1}{\pi} \int_0^\pi e^{2r\cos \theta} \,d\theta \leq e^{2|r|},
$$
or by brute force.
\end{proof}

The one-half power appearing in \eqref{ab_bound} is the smallest possible; see Example~\ref{E2} below.

In the proof of Theorem~\ref{T1} will specifically consider only $\lambda\in\Reals$ in order to be able
to use some facts about self-adjoint operators.
When $u_0$ is real-valued, Theorem~\ref{T1} follows by the argument presented in Lemma~3 of \cite{S}.
Of course the problem is unchanged if $u_0$ is a complex multiple of a real solution.  However, when
$\Re u_0$ and $\Im u_0$ are linearly independent solutions, one needs to make some modifications.  The
approach we take is to show that one may replace $u_0$ by $\Re u_0$:

\begin{prop}\label{P2}
Suppose $\Re u_0$ and $\Im u_0$ are linearly independent and $b(\lambda)\equiv 0$.  Then
$W[\Re u_0,\Re u_\lambda](x) = 0$ for all $x>1$ and all $\lambda\in\Reals$.
\end{prop}

\begin{proof}
By assumption, $u_\lambda(x)=a(\lambda)u_0(x)$ for all $x>1$.  For $\lambda\in\Reals$, both $u_\lambda$ and
$\bar u_\lambda$ are solutions of \eqref{Eq2}.  Therefore,
$$
W[u_0,\bar u_0] = W[u_\lambda,\bar u_\lambda] = |a(\lambda)|^2 W[u_0,\bar u_0],
$$
which is non-zero because we assumed linear independence.  Thus it follows that $a(\lambda)$ is unimodular
for $\lambda\in\Reals$.

By the Schwarz reflection principle, the complex conjugate of any zero of $a$ must be a pole; however, $a$ is
an entire function so we may conclude that it is zero-free.  This means that $\log[a(\lambda)]$ is an entire
function, but then by the estimate in Proposition~\ref{P1}, it must be constant.  By taking $\lambda=0$, we
learn that $a(\lambda)\equiv 1$.

We have just seen that for all $\lambda$ (real or complex) and all $x>1$, $u_\lambda(x)=u_0(x)$.
By taking real parts, we immediately obtain the conclusion sought.
\end{proof}

\begin{proof}[Proof of Therorem~\ref{T1}]
In light of Proposition~\ref{P2} we may assume that $u_0$ is real-valued.  We now essentially repeat the argument from
Lemma~3 of \cite{S}.  Let us choose $\theta_0$ and $\theta_1$ so that
$$
\cos(\theta_0) u_0(0) + \sin(\theta_0) u_0'(0) = 0 = \cos(\theta_1) u_0(1) + \sin(\theta_1) u_0'(1)
$$
and consider the self-adjoint operators $H_\lambda u = -u''+Qu+\lambda Vu$ on $[0,1]$
with the boundary conditions
$$
\cos(\theta_x) u(x) + \sin(\theta_x) u'(x) = 0 \quad\text{for $x\in\{0,1\}$.}
$$

Suppose $V \neq 0$ and the set of $\lambda$ for which $b( \lambda) =
0$ has an accumulation point. In this case, $b( \lambda) \equiv 0$,
and in particular, $0$ is an eigenvalue of $H_\lambda$ for every $\lambda\in\Reals$.  As the spectrum of $H_\lambda$ is
simple, discrete, and bounded from below, this implies that the number of negative eigenvalues of $H_\lambda$
is finite and independent of $\lambda$.  We will derive a contradiction by using a very weak form of Weyl's Law.
(When $Q$ and $V$ obey some mild regularity hypotheses, full Weyl-Law asymptotics are known,
cf. \cite[Theorem~XIII.79]{RS4}.)

As $V\neq0$, it must happen that either $V$ is positive on a set of positive measure, or $V$ is negative on a set
of positive measure.  We will treat the former case, the latter follows with obvious modifications.

For each $\epsilon>0$, let us define
$$
\phi_\epsilon(x)= \begin{cases}\sqrt{\tfrac3{2}} \epsilon^{-3/2} (\epsilon - |x|) &\ |x|<\epsilon \\
0 &\ |x|\geq \epsilon \end{cases}
$$
By the Lebesgue differentiation theorem, $\int\phi(x-t)^2 V(t)\,dt \to V(x)$ as $\epsilon\downarrow0$
for a.e. $x\in\Reals$.  Therefore, given any integer $N>0$, one can find $N$ distinct points
$x_1,\ldots,x_N\in(0,1)$ and $\epsilon$ sufficiently small, so that $\phi_\epsilon(x-x_j)$ are supported
in disjoint subsets of $(0,1)$ and obey $\int V(x) \phi_\epsilon(x-x_j)\,dx > \epsilon$.  Thus, by the minimax
principle (cf. \cite[\S{}XIII.1]{RS4}) one can see that $H_\lambda$ has at least $N$ negative eigenvalues
when $\lambda$ is a sufficiently large negative number.
\end{proof}

An alternate approach to proving that $a(\lambda)\equiv 1$ implies $V\equiv 0$, is through
recent work on the large-$\lambda$ asymptotics of the spectral shift function $\arg[a(\lambda)]$; see
\cite{Push,Safr}, for example.  We would like to thank A.~Pushnitski for explaining some of this material to us.

\begin{proof}[Proof of Therorem~\ref{T2}]
The conclusion of this theorem holds for any non-zero entire function of order one-half and finite type;
see \cite[Theorem~2.5.13]{Boas} or \cite[Theorem~5.2.1]{SteinShak}.
\end{proof}

Therorem~\ref{T2} is optimal with regard to the power of $r$.  This is to be expected from Weyl's Law
and can be seen with an elementary example:

\begin{Example}\label{E2}
Consider $Q\equiv-\pi^2$, $u_0(x)=\sin(\pi x)$, and $V=-\chi_{[0,1]}$. In this case,
$$
b(\lambda) = -\cos\Bigl(\sqrt{\lambda+\pi^2}\, \Bigr).
$$
As cosine is an even function, both branches of the square-root lead to the same answer.
\end{Example}


\begin{thebibliography}{99}

\bibitem{Boas} R. P. Boas Jr., \textit{Entire functions.}
Academic Press Inc., New York, 1954.

\bibitem{Chadan} K.~Chadan,
Le probl\`{e}me inverse en constante de couplage.
\textit{C. R. Acad. Sci. Paris S\'er. II M\'ec. Phys. Chim. Sci. Univers Sci. Terre} \textbf{299} (1984), 271--274.

\bibitem{DLS} D. Damanik, D. Lenz, and G. Stolz
Lower Transport Bounds for One-Dimensional Continuum Schr\"odinger Operators.
\texttt{arXiv:math-ph/0410062}.

\bibitem{DSS} D. Damanik, R. Sims, and G. Stolz, Localization for
one-dimensional, continuum, Bernoulli-Anderson models. {\it Duke
Math. J.} {\bf 114} No. 1 (2002) 59--100.

\bibitem{DSS2} D. Damanik, R. Sims, and G. Stolz, Localization for
discrete one-dimensional random word models. \textit{J. Funct. Anal.}
\textbf{208} (2004), 423--445.

\bibitem{DatKun} P. K. Datta and K. Kundu, The absence of localization in
one-dimensional disordered harmonic chains. \textit{J. Phys.
Condens. Matter.} {\bf 6} (1994), 4465--4478.

\bibitem{BiGer} S. de Bi\`{e}vre and F. Germinet, Dynamical
localization for the random dimer Schr\"{o}dinger operator,
\textit{J.\ Stat.\ Phys.} {\bf 98} (2000), 1135--1148.

\bibitem{JSS} S. Jitomirskaya, H. Schulz-Baldes, G. Stolz,
Delocalization in random polymer models.
{\em Commun. Math. Phys.},  {\bf 233} (2003), 27--48.

\bibitem{Push} A.~Pushnitski,
Spectral shift function of the Schr\"odinger operator in the large coupling constant limit.
Comm. Partial Differential Equations \textbf{25} (2000), 703--736.

\bibitem{RS4} M. Reed and B. Simon,
\textit{Methods of modern mathematical physics. IV. Analysis of operators.}
Academic Press, New York-London, 1978.

\bibitem{Safr} O.~Safronov,
Spectral shift function in the large coupling constant limit.
\textit{J. Funct. Anal.} \textbf{182} (2001), 151--169.

\bibitem{SteinShak} E. M. Stein and R. Shakarchi,
\textit{Complex analysis.}
Princeton Lectures in Analysis, II.
Princeton University Press, Princeton, NJ, 2003.

\bibitem{S} G. Stolz, Non-monotonic Random
Schr\"odinger Operators: The Anderson Model.
\textit{J. Math. Anal. Appl.} {\bf 248} (2000), 173--183.

\bibitem{Yoshida} K.~Yoshida,
\textit{Lectures on differential and integral equations.}
Interscience Publishers, New York, 1960.
\end{thebibliography}
\end{document}